\documentclass{Rinton-P9x6}

\newcommand{\beq}{\begin{equation}}
\newcommand{\eeq}{\end{equation}}
\newcommand{\beqa}{\begin{eqnarray}}
\newcommand{\eeqa}{\end{eqnarray}}
\newcommand{\ket} [1] {\vert #1 \rangle}

\begin{document}

\title{Generation of Large Photon-Number Cat States using Linear Optics and Quantum Memory\thanks{\rm \tiny To appear in the Proceedings of the Sixth International Conference on Quantum Communication, Measurement and Computing (QCMC'02), Boston, July 22-26, 2002.}}

\author{N. J. Cerf$^{1}$, J. Fiur\'{a}\v{s}ek$^{2}$, S. Iblisdir$^{1}$, and  S. Massar$^{1,3}$}
\address{$^{1}$ Ecole Polytechnique,
CP 165, Universit\'{e} Libre de Bruxelles, B-1050 Bruxelles, Belgium \\
$^{2}$Department of Optics, Palack\'{y} University, 17. listopadu 50,
77200 Olomouc, Czech Republic \\
$^{3}$Service de Physique  Th\'{e}orique, CP 225,
Universit\'{e} Libre de Bruxelles, B-1050~Bruxelles, Belgium}

\maketitle

\abstracts{A recursive method for producing path-entangled states of light 
is presented. These states may find applications in quantum lithography and high-precision interferometric measurements. The required resources are single-photon sources, linear optics components, and photodetectors. Adding a quantum memory greatly enhances the yield in comparison with the previously known schemes.}

Entangled states play a fundamental role in quantum information processing as they are basic ingredients in various tasks such as quantum teleportation, quantum key distribution, or quantum computing \cite{nielsen}. Entanglement has also been shown to be a useful resource for high-precision frequency measurements or quantum lithography \cite{boll96,boto00}. In this paper, we will focus on the entangled states needed in the latter applications, that is, photon-number path-entangled states or \emph{cat states}. These states are of the form
\beq\label{eq:cat}
\ket{N_+}_{a,b}=\frac{1}{\sqrt{2 N!}}(a^{*N}+b^{*N})\ket{0},
\eeq 
where $a^*$ and $b^*$ denote the bosonic creation operators for two modes of the electromagnetic field, and $\ket{0}$ denotes the vacuum state. 
Deterministic schemes to produce cat states using $\chi^{(3)}$ nonlinearity have been proposed \cite{milb89,gerr01}. Unfortunately, the presently available $\chi^{(3)}$ couplings are about $10^{16}$ times weaker than the required values, which makes these schemes out of the scope of current technology. Recently, however, alternative schemes that only require photon-number state sources, linear optics components (beam splitters and phase shifters) and photodetectors have been proposed for $N=4$ first\cite{lee02}, and then for an arbitrary photon number $N$\cite{kok02,fiur02:noon,zou02}. These schemes are unfortunately \emph{probabilistic}, and the probability $p(N_+)$ of a successful state preparation scales exponentially poorly with $N$ for all of these schemes. Typically, if $N$ is even, $p(N_+)$ scales as $c^{-N}$ with the constant $c=\sqrt{2}\; e$ for the scheme \cite{kok02} and $c=e$ in \cite{fiur02:noon}. This is so because these schemes work by adding (or subtracting) the photons one by one (or two by two), each step having a non-unity probability of success. Such a multiplicative process (the yield is a product of success probabilities) naturally scales exponentially.
\par

In this paper, we will demonstrate that using a \emph{tree-like} structure for the linear optics circuit makes it possible to significantly enhance the yield. We will describe a recursive scheme that generates a state of the form of Eq.~(\ref{eq:cat}) when $N$ is a power of $2$ with an approximate yield of
\beq\label{eq:mainresult}
(4\sqrt{\pi})^{-\log_2 N} N^{(1-\log_2 N)/4}.
\eeq
Thus, in contrast with the previous schemes, the scaling is \emph{sub-exponential} in $N$, which exploits the fact that only
$\log_2 N$ recurrence levels are needed in the tree. As we will see, the price to pay is that a quantum memory (or fast optical switches) is required in addition to the other optical components.
\par

The basic operation in our method is the transformation $T_N$, which probabilistically transforms two cat states of $N$ photons into one cat state containing $2N$ photons, that is, $T_N: \ket{N_+}_{a,b}\otimes  \ket{N_+}_{c,d}\rightarrow \ket{2N_+}_{a,b}$ with probability $p(T_N)$. 
The transformation $T_N$ is realized as follows. First, change the phase of mode $d$ as $d \to e^{i \pi /N} d$. Next, combine the modes $a$ and $c$ at a $50$:$50$ beam splitter and modes $b$ and $d$ at another $50$:$50$ beam splitter. Defining the action of a $50$:$50$ beam splitter on two impinging modes $a$ and $b$ as $a \rightarrow (a-ib)/\sqrt{2}$ and $b \rightarrow (-ia+b)/\sqrt{2}$, we obtain the 4-mode state
\beq\label{eq:tn} 
{1 \over 2 ^{N+1}  N! }\left((a^* + i c^*)^N + (b^* + i d^*)^N \right) 
\left ((ia^* +  c^*)^N - (ib^* +  d^*)^N \right) |0\rangle.
\eeq
Finally, measure the number of photons contained in modes $c$ and $d$. 
The transformation succeeds if \emph{no} photon is detected in these two modes, in which case we get the (unnormalized) state 
\beq\label{eq:tn2}
{1 \over 2 ^{N+1}  N! }
 \left(
(a^*)^{2N} -
(b^*)^{2N} \right) 
 |0\rangle .
\eeq
Transforming the phase of mode $b$ as $b \to e^{i \pi /2N} b$ then yields the desired state $\ket{2N_+}_{a,b}$ with a success probability  $p(T_N)={(2N)! \over 2^{2N+1} (N!)^2}$. Upon using the Stirling formula, we get 
\beq  \label{eq:stirling}
p(T_N)\simeq {1 \over \sqrt{4\pi N}} \left(1+O(1/N)\right).
\eeq
Remarkably, $p(T_N)$ does not decrease exponentially with $N$ as if the photons had been added one by one, but only as $1/\sqrt{N}$, so the successive steps of our iterative process keep a reasonably good efficiency as $N$ increases.
\par

Consider first a naive scheme that does not use a quantum memory. Starting from $N$ one-photon states $\ket{1_+}$, which are easily obtainable with a single-photon source and a beam splitter, one probabilistically produces $N/2$ states $\ket{2_+}$, then $N/4$ states $\ket{4_+}$, and so on until one gets the desired state $\ket{N_+}$. What is the overall probability $p(N_+)$ to produce $\ket{N_+}$ with this procedure? Solving the recurrence equation $p(2N_+)=p(N_+)^2 p(T_N)$, we get
\beq
p(N_+) = {2 N! \over (2N)^N} \simeq \sqrt{8\pi N} \; (2e)^{-N} \left(1+O(1/N)\right).
\eeq
The reason why this success probability decreases exponentially with $N$ is that we require all the $(N-1)$ individual transformations $T$ at all recurrence levels to succeed simultaneously. Since each transformation $T$ has a non-unity probability of success, the probability that they all succeed is exponentially small. 
\par

Let us now show that we can strongly improve on this by using a quantum memory. We start with $\ket{1_+}^{\otimes M_1}$ where the number $M_1$ of states we need to prepare initially will be determined below. Suppose that, after a few steps, we have obtained the state $|n_+\rangle^{\otimes M_n}$, assuming $M_n$ is an even number. We carry out the transformation $T_n$ on each pair of states, which produces on average $M_{2n}$ states $|2n_+\rangle$ where $M_{2n} = {M_n \over 2} P(T_n).$
We then discard the states for which the transformation failed and only keep those for which it succeeded (there are, on average, $M_{2n}$ such states). We use these states to construct the state $|4n_+\rangle$, and so on. If we want to produce $M_N$ cat states $|N_+\rangle$ (where $N$ is a power of 2) with a probability of order one, then, solving the above recurrence using Eq.~(\ref{eq:stirling}), we find that the average number $M_1$ of states $\ket{1_+}$ that are needed initially is
\begin{equation}
M_1 \simeq (4 \sqrt{\pi})^{\log_2 N} N^{(\log_2 N - 1)/4} 
\left(M_N+{\cal O}(\sqrt{M_N})\right) ,
\end{equation}
which immediately implies Eq.~(\ref{eq:mainresult}). 
This shows that the resources ({\it i.e.}, the number of single photons, of operations $T_n$, of beam splitters, of photodetections, etc.) needed for generating a cat state increase only sub-exponentially with $N$.
This recursive scheme thus requires much less
operations than the naive scheme, but the dynamics is now conditional on whether the operations $T_n$ succeed or not. Such a conditional dynamics requires fast optical switches\cite{kwia99} in order to store the photons in a loop, or, alternatively, a quantum memory\cite{scho02} in which the photon states can be stored. These extra technological resources make our scheme much more demanding than those based only on linear optics, single photon sources, and photodetectors only \cite{lee02,kok02,fiur02:noon,zou02}, but $\chi^{(3)}$ couplings are still not required. Note that we may start our recursion from two-photon cat states $\ket{2_+}$ (instead of $\ket{1_+}$) since they can be deterministically generated from two one-photon states using a beam splitter\cite{hong87}, resulting in an extra factor of 4 in the yield Eq.~(\ref{eq:mainresult}).
\par

Our scheme is sensitive to various imperfections such as the losses in the beam splitters, imperfect mode-matching, etc. Let us focus on one of them, namely the non-unity detector efficiency $\eta < 1$. The probability that the detector does not click when it has absorbed $j$ photons is given by $(1-\eta)^j$. At first sight, it seems that the main source of errors will be due to the events where no photon is detected in modes $c$ and $d$ while a photon has actually been absorbed in either one of the detectors. By expanding Eq.~(\ref{eq:tn}) in a series in $c^*$ and $d^*$, we find that the probability that there is a single photon in mode $c$ or $d$ is $p(c^*c+d^*d=1)={2 N \over 2^{2N}}$. This probability is thus exponentially smaller than $p(T_N)$, so this event can in fact be neglected in good approximation. Next, if we calculate the probability that two photons are in mode $c$ or $d$, $p(c^*c+d^*d=2)={(2N-2)! \over 2^{2N +1} (N-1)!^2}\simeq {P(T_N) \over 4}$, we find that, for large $N$, the probability that one decides that $T_N$ succeeded while it actually failed is approximately given by $(1 - \eta)^2 /4$. So we conclude that our scheme is robust with respect to detector inefficiencies
up to the second order in $(1-\eta)$.
\par

We thank Stefano Pironio and JiangFeng Du for contributing to stimulating discussions and the European Science Foundation for its financial support. 
S.I. acknowledges support from the FRIA foundation (Belgium). S.M. is Research Associate of the FNRS (Belgium). This work was partly funded by the European Union under the project EQUIP (IST-FET programme).

\end{document}